\documentclass[iop]{emulateapj}

\newcommand{\be}{\begin{equation}}
\newcommand{\ee}{\end{equation}} 

\shorttitle{High-$z$ Quasars}
\shortauthors{Melia}

\begin{document}

\title{High-$z$ Quasars in the $R_{\rm h}=ct$ Universe}
\author{Fulvio Melia$^\dag$\\
Key Laboratory of Dark Matter and Space Astronomy, Purple Mountain Observatory,\\ 
Chinese Academy of Sciences, 210008 Nanjing, China\\
and\\
Department of Physics, The Applied Math Program, and Department of Astronomy,\\  
The University of Arizona, AZ 85721, USA\\
$^\dag${John Woodruff Simpson Fellow. E-mail: melia@as.arizona.edu}}

\begin{abstract}
One cannot understand the early appearance of $10^9\,M_\odot$ supermassive black holes
without invoking anomalously high accretion rates or the creation of exotically massive seeds,
neither of which is seen in the local Universe. Recent observations have compounded this
problem by demonstrating that most, if not all, of the high-$z$ quasars appear to be accreting
at the Eddington limit. In the context of $\Lambda$CDM, the only viable alternative now
appears to be the assemblage of supermassive black holes via mergers, as long as the seeds
started forming at redshifts $>40$, but ceased being created by $z\sim 20-30$.
In this paper, we show that, whereas the high-$z$ quasars may be difficult to explain
within the framework of the standard model, they can instead be interpreted much more sensibly
in the context of the $R_{\rm h}=ct$ Universe. In this cosmology, $5-20\;M_\odot$ seeds
produced {\it after} the onset of re-ionization (at $z\la 15$) could have easily grown to 
$M\ga 10^9\;M_\odot$ by $z\ga 6$, merely by accreting at the standard Eddington rate.
\end{abstract}

\keywords{cosmology: observations, theory; dark ages; reionization; early universe; quasars: general}

\section{Introduction}
Quasars are the most powerful objects in the Universe. The efficient extraction
of gravitational energy from matter falling onto their central supermassive 
black hole is sufficient to make them identifiable out to very large redshifts ($z\ga7$), 
where they can be studied to probe black-hole growth (e.g., Kauffmann \& Haehnelt 
2000; Wyithe \& Loeb 2003; Hopkins et al. 2005; Croton et al. 2006) and the 
formation of large-scale structure in the early Universe (Fan 2006). For example, 
it is now believed that supermassive black holes played an active role in galaxy 
evolution, e.g., through their feedback on the interstellar medium, by heating 
and expelling gas that would otherwise have formed stars (see also Silk \& Rees 
1998; Di Matteo et al. 2005). However, the impact of studying high-$z$ quasars is not
limited to just these few examples; they can also provide crucial information on the 
re-ionization of the intergalactic medium (IGM) several hundred Myr after recombination.

One of the most useful observations of these sources has been the reverberation
mapping of broad-lines, yielding the distance of the line-emitting gas from the
central ionizing source. The measurement of the velocity of this plasma, e.g.,
via its Doppler-broadened line width, can therefore yield the gravitational
mass of the central black hole to within a factor 3 or so (e.g., 
Wandel et al. 1999). This method has made it possible to determine black-hole
masses for quasars beyond $z\sim 6$ (Willott et al. 2003; Jiang et al.
2007; Kurk et al. 2007; Kurk et al. 2009) and, combined with a measurement of
the source spectrum, has helped us conclude that the most luminous 
high-$z$ quasars contain black holes with mass $M\ga 10^9\;M_\odot$, 
accreting at close to the Eddington limit.

But the discovery of such massive objects so early in the Universe's history
(corresponding to an age of only $\sim 800$ Myr in the context of $\Lambda$CDM),
and the apparent constancy of the observed quasar broad-line region properties
with redshift (e.g., Pentericci et al. 2002), is quite surprising and difficult to
understand in the context of the standard model. Indeed, the 
early appearance of supermassive black holes has developed into 
one of the most important unsolved mysteries in astronomy.

Their existence has resulted in diverse attempts to 
explain how such objects could have formed so quickly in the rapidly
expanding early Universe. If one invokes standard Eddington-limited
accretion, the e-folding time (i.e., the so-called Salpeter timescale)
for black-hole growth is $\sim 45$ Myr, assuming an efficiency of $10\,\%$.
Thus, reaching an observed mass of $10^9\,M_\odot$ from a supernova-produced
seed black hole of $5\;M_\odot$ would have taken at least 860 Myr, longer than the
short cosmic time available since the big bang. 

Some workers have therefore suggested that super-Eddington
accretion must have occurred at very high redshifts (e.g., Volonteri \& Rees 2005). 
Others have proposed that rapidly spinning black holes accrete with higher efficiency 
(when the disk is spinning with a retrograde orbit), thereby lowering the Salpeter 
timescale (Shapiro 2005). It has also been demonstrated that Eddington-limited
accretion can still account for these black holes, but only if the seeds
were $>> 5\;M_\odot$. However, to produce 
high-$z$ quasars in only $\sim 400$ Myr, the initial black hole 
masses would have had to have been greater than $\sim 10^{4-5}\;M_\odot$ 
(Yoo \& Miralda-Escud\'e 2004). 

All the proposals thus far appear to be contrived for one 
reason or another. Unless some exotic mechanism involving self-interacting
dark matter was responsible, the resolution to this problem 
ultimately relies on an enhanced accretion rate, or the creation of a
seed mass

{\baselineskip 12pt
\begin{deluxetable*}{lccccrrl}
\tabletypesize{\scriptsize}
\tablecolumns{8}
\tablewidth{0pc}
\tablecaption{High-$z$ Quasars}
\tablehead{
\colhead{Name} & \colhead{$\quad z\quad$} &$M/10^8\,M_\odot$& \colhead{Age/Myr}    
& \colhead{Age/Myr} 
& \colhead{Seed/Myr} &\colhead{Seed/Myr}&\colhead{Ref.} \\
&  &  & \colhead{($\Lambda$CDM)}    
& \colhead{($R_{\rm h}=ct$)} 
&($5\,M_\odot$) & ($20\,M_\odot$)&
}
\startdata
ULAS J1120+0641& 7.085 &$\sim$20&737& 1,634  &743$\;$&805$\;\;$&Mortlock et al. (2011)\\
CFHQS J0210-0456& 6.440 &0.4--1.4&834 & 1,776 &1,004$\;$&1,123$\;\;$&Willott et al. (2010a)\\
CFHQS J2329-0301& 6.430 & $\sim$2 &836  & 1,778  &990$\;$ &1,052$\;\;$&Willott et al. (2007)\\
SDSS J1148+5251& 6.419 &20--123&838  & 1,781 &808$\;$&952$\;\;$&Fan et al. (2002)\\
& && &&&&De Rosa et al. (2011)\\
CFHQS J2329-0301& 6.417 &2.1--2.9&838 & 1,781 &977$\;$&1,053$\;\;$&Willott et al. (2010a)\\
SDSS J1030+0524& 6.280 &8--12 &862  & 1,815 &947$\;$&1,027$\;\;$&Pentericci et al. (2002)\\
CFHQS J0050+3445& 6.253 &22--31&867 & 1,822 &911$\;$&989$\;\;$&Willott et al. (2010b)\\
QSO J1623+3112& 6.220 &12--18&874 & 1,830 &943$\;$&1,024$\;\;$&Wang et al. (2007)\\
SDSS J1048 + 4637&6.198&16--98&877&1,835&872$\;$&1,016$\;\;$&De Rosa et al. (2011)\\
CFHQS J0221-0802& 6.161 &2.3--14.5&885 & 1,845 &968$\;$&1,113$\;\;$&Willott et al. (2010b)\\
CFHQS J2229+1457& 6.152 &0.7--1.9&886 & 1,848 &1,063$\;$&1,170$\;\;$&Willott et al. (2010b)\\
CFHQS J1509-1749& 6.121 &27--33&892 & 1,855 &941$\;$&1,013$\;\;$&Willott et al. (2007)\\
CFHQS J2100-1715& 6.087 &9.2--12.3&897 & 1,863 &995$\;$&1,069$\;\;$&Willott et al. (2010b)\\
SDSS J0303-0019& 6.080 &1.5--3.8& 899& 1,866 &1,049$\;$&1,154$\;\;$&Kurk et al. (2009)\\
SDSS J0353 + 0104&6.072&6--35&900&1,868&952$\;$&1,093$\;\;$&De Rosa et al. (2011)\\
SDSS J0842 + 1218&6.069 &7--43&901&1,869&943$\;$&1,025$\;\;$&De Rosa et al. (2011)\\
SDSS J1630 + 4012&6.058&3.6--22.5&903&1,872&975$\;$&1,120$\;\;$&De Rosa et al. (2011)\\
CFHQS J1641+3755& 6.047 &1.6--3.4&906 & 1,876 &1,064$\;$&1,161$\;\;$&Willott et al. (2007)\\
SDSS J1306 + 0356&6.018&7--43&911&1,882&956$\;$&1,100$\;\;$&De Rosa et al. (2011)\\
SDSS J1308+0356& 6.016 &10--12&911 & 1,883 &1,015$\;$&1,085$\;\;$&Jiang et al. (2007)\\
SDSS J1306+0356& 5.990 &2.0--2.8&916 & 1,890 &1,087$\;$&1,165$\;\;$&Jiang et al. (2007)\\
CFHQS J0055+0146& 5.983 &1.7--3.3&917 & 1,892 &1,082$\;$&1,174$\;\;$&Willott et al. (2010a)\\

\enddata
\end{deluxetable*}
}

\noindent much larger than we can currently explain. Observations of 
the local Universe indicate that seed black holes could have formed
during supernova explosions following the Dark Ages (see \S~2 below), 
but probably not with masses exceeding $\sim 5-20\;M_\odot$. And
now that we have detected over 50 high-$z$ quasars, we also know that
none of them appear to be accreting at greatly super-Eddington rates
(see, e.g., figure~5 in Willott et al. 2010a). 

In the context of the standard model, the only viable alternative now appears 
to be the role played by mergers in the early Universe, which may also be
relevant---perhaps even essential---to creating the correlation
between black-hole mass and host bulge mass observed nearby
(see, e.g., Tanaka \& Haiman 2009; Lippai et al. 2009; Hirschmann
et al. 2010). However, in order to make this work, the black-hole seeds
would have had to form well before the currently understood beginning
of the period of reionization, and their creation would have had to cease
by a redshift of $\sim 20-30$ in order to not overproduce the mass density
in lower-mass (i.e., between a few $\times 10^5\;M_\odot$ and
a few $\times 10^7\;M_\odot$) black holes. We will continue our
discussion of this possibility in the Conclusions section.

In this paper, we advance an alternative viewpoint---that the difficulty
faced by $\Lambda$CDM in accounting for the high-$z$ quasars
is actually more evidence against it being the correct cosmology. We 
show that the problems faced by $\Lambda$CDM disappear when we 
instead interpret the origin and evolution of the high-$z$ quasars in the 
context of the $R_{\rm h}=ct$ Universe. In particular, we will demonstrate that in this 
cosmology, high-$z$ quasars appeared only after the beginning of the 
Epoch of Re-ionization (EoR), as one would expect if their 
$5-20\;M_\odot$ seeds were created from the deaths of Pop 
II and III stars.

\vskip 0.2in
\section{The Dark Ages and Epoch of Re-ionization}
The explanation for how and when the high-$z$ quasars came into existence is
necessarily woven into the history of  the Dark Ages and the EoR
that followed. In the context of $\Lambda$CDM, the Universe became transparent
roughly 0.4 Myr after the big bang, ushering in a period of darkness that ended several
hundred Myr later, when the first stars and galaxies started forming and emitting ionizing
radiation. The current constraints delimit the EoR to the redshift range
$z\sim 6-15$ (Zaroubi 2012), which (again in the context of $\Lambda$CDM) 
corresponds to a cosmic time $t\sim 400-900$ Myr.   

The IGM was neutral at the beginning of the EoR, but an evolving patchwork of
neutral (HI) and ionized hydrogen (HII) unfolded as the re-ionization progressed.
The temperature and ionized fraction of the gas increased rapidly as the number 
of ultraviolet sources formed, until the ionized regions eventually permeated the
whole Universe (Barkana \& Loeb 2001; Bromm \& Larson 2004). Some of the best
probes of this re-ionization process are actually the high-$z$ quasars themselves.
High-resolution spectra of SDSS high-$z$ quasars show a complete absence
of structure bluewards of the quasar Lyman-$\alpha$ restframe emission, especially
those with redshift $\ga 6$ (Fan et al. 2006). This is interpreted as due to an increase
in the Lyman-$\alpha$ optical depth resulting from the decrease in the ionized
fraction along the line-of-sight. The fact that the trend is not monotonic with
redshift, may also be an indication that we are witnessing the aforementioned
patchwork of ionization at these redshifts. The main conclusion
from the Lyman-$\alpha$ optical depth measurements is that the Universe
was highly ionized at $z\la 6$, but its neutral fraction increased with
increasing redshift, presumably reaching a value $\sim$1 by $z\sim 15$.
Support for these results is also provided by the Wilkinson Microwave Anisotropy
Probe (WMAP, Bennett et al. 2003) mission, whose measurements suggest that
the Universe may have been $\sim 50\%$ neutral at $z\ga 10$, and
that re-ionization did not start before $z\sim 14$ (Page et al. 2007).

Hydrogen is ionized by photons with energy $\ga 13.6$ eV. Obvious candidate
sources are Pop II and III stars. The constraints on decaying or self-annihilating
dark matter particles, or decaying cosmic strings, make it unlikely that these
could re-ionize the Universe on their own (see, e.g., Ripamonti et al. 2007).
Accreting supermassive black holes also produce large quantities of UV and 
X-radiation, but their mass distribution in the early Universe is not fully
known, rendering the role played by quasars during re-ionization somewhat
uncertain (see, e.g., Meiksin 2005). We can therefore reasonably conclude
that the EoR was brought about by a combination of Pop II and III
stars in star-forming galaxies and the ionizing radiation produced by the
high-$z$ quasars that formed during this period. The latter may have
become more important towards the end of the EoR, though it is not
clear if they, by themselves, could have caused the Universe to become
fully ionized by redshift $z\la 6$ (Jiang et al. 2006).

Let us now examine how the high-$z$ quasars observed thus far impact this 
basic picture when viewed from the perspective of $\Lambda$CDM. In Table 1, 
we list many of the objects for which a reliable determination of mass has
been made. This is not a complete sample, but is probably quite representative
of the overall distribution. The columns in this table are as follows: (2) the
quasar's measured redshift, (3) its inferred mass, quoted as a range when
the uncertainty is well defined, (4) the quasar's age, calculated from its
redshift, using the Universe's expansion history according to $\Lambda$CDM,
(5) the quasar's age (for the same redshift) calculated using $R_{\rm h}
=ct$, (6) the cosmic time at which a $5\;M_\odot$ seed must have
formed in order to grow into the observed quasar, assuming Eddington-limited
accretion in the context of $R_{\rm h}=ct$, and (7) the same as column
(6), except for a seed mass of $20\;M_\odot$. 

Our analysis in this paper is based strictly on the premise that black-hole
seeds are produced in supernova events, which observations in the local
Universe indicate result in $\sim5-20\;M_\odot$ remnant cores. We further
avoid any possible exotic mechanism that might have enhanced their
accretion rate above the Eddington limit, defined from the maximum
luminosity attainable due to outward radiation pressure acting on highly
ionized infalling material (see, e.g., Melia 2009). This power depends
somewhat on the gas composition, but for hydrogen plasma is given as
$L_{\rm Edd}\approx 1.3\times 10^{38}(M/M_\odot)$ ergs s$^{-1}$,
in terms of the accretor's mass, $M$. The accretion rate $\dot{M}$ is inferred
from $L_{\rm Edd}$, once the fractional efficiency $\eta$ of converting rest mass
energy into radiation is identified. Matter spiraling inwards towards
the event horizon releases a minimum of $\sim6\%$ of its rest energy
for a Schwarzschild black hole, but $\eta$ may be as high as $0.3$
(or more) when the black hole is spinning (and the disk is on a prograde
orbit), permitting frictional dissipation
to continue down to radii smaller than $\sim 3$ Schwarzschild radii
(e.g., Melia 2009). Most workers adopt the ``fiducial" value $\eta=0.1$,
reasonably close to the Schwarzschild limit, but also allowing for some
influence from a possible black-hole spin. This is the value we will
assume here.

For simplicity, we will also ignore the contribution to black-hole growth from mergers, which
almost certainly must have occurred at some level and, as we have already indicated in
the introduction, may have been necessary in order to produce their final distribution and
intrinsic scatter in black-hole versus galaxy mass relations (Hirschmann et al. 2010). The
constraints we will derive below should therefore be viewed with this caveat
in mind. 

Under these conditions, the maximum accretion rate experienced by
quasars at high redshift was then $\dot{M}=L_{\rm Edd}/\eta c^2$, which
grew their mass at a rate $\dot{M}\approx 0.022\,(M/M_\odot)\;M_\odot/$Myr.
Solving for the mass as a function of cosmic time then produces the Salpeter
relation, 
\begin{equation}
M(t)=M_0 \exp\left({t-t_{\rm seed}\over 45\;{\rm Myr}}\right)\;,
\end{equation}
where $M_0$ is the seed mass produced at time $t_{\rm seed}$. We have calculated
the time $t_{\rm seed}$ for each of the quasars in Table 1, using the Universe's
expansion history in $\Lambda$CDM, for two values of the seed mass,
$M_0=5\;M_\odot$ and $20\;M_\odot$. In each case, the possible range of ``birth"
times (between these two limits) is shown against the redshift
of the observed quasar in figure~1, for the assumed cosmological parameters
$H_0=72$ km s$^{-1}$ Mpc$^{-1}$, $\Omega_m=0.28$, and a dark-energy 
equation of state $p=-\rho$.

In this figure, we also show the period of Dark Ages and the EoR, the latter identified
from its measured redshift range ($z\sim 6-15$). The evident problems with this diagram are well
known by now and we do not need to dwell on them. We simply point out the rather
obvious flaws with this model, in which all of the quasars in this sample had to begin
their evolution from seeds formed around the time of the big bang, well before the EoR even
started. Worse, as quasars continue to be discovered at greater and greater redshifts, 
the model becomes completely untenable. Already, ULAS J1120+0641 (discovered at 
redshift 7.085; Mortlock et al. 2011), and SDSS J1148+5251 (discovered at
redshift 6.419; Fan et al. 2002),  must both have started their growth phase well
before the big bang, which makes no physical sense. 

These are the reasons, of course, why investigators have been motivated to
propose exotic methods of greatly enhancing
the accretion rate and/or starting from much bigger seed masses. But as we have
already suggested in the introduction, our proposal in this paper is that the
black-hole birth and evolution were in fact quite standard (as we understand these
from observations in the local Universe), and that the answer lies instead with the
assumed cosmology. Next, we will move to a consideration of how the observations
are interpreted in the context of the $R_{\rm h}=ct$ Universe. 

\section{The $R_{\rm h}=ct$ Universe}
Whereas $\Lambda$CDM is an empirical cosmology, deriving many of its traits from
observations, the $R_{\rm h}=ct$ Universe is an FRW cosmology whose basic
principles follow directly from a strict adherence to the Cosmological 
Principle and Weyl's postulate (Melia 2007; Melia \& Shevchuk 2012; see also
Melia 2012a for a more pedagogical treatment). By now, these two expansion
scenarios have been compared with each other, and with the various
cosmological measurements, both at low and high redshifts. The evidence
suggests that $R_{\rm h}=ct$ is a better match to the data, particularly
the CMB, whose fluctuations are so different from the predictions of 
$\Lambda$CDM that its probability of being the correct description of
nature has been assessed at $<0.03\%$ (Copi et al. 2009). On the other
hand, the $R_{\rm h}=ct$ Universe explains the CMB angular correlation
function very well, particularly the observed absence of any correlation
at angles greater than $\sim 60^\circ$ (Melia 2012b). A short summary
of the current status of this cosmology appears in Melia (2012c).

\begin{figure}[hp]
{\centerline{\epsscale{1.1} \plotone{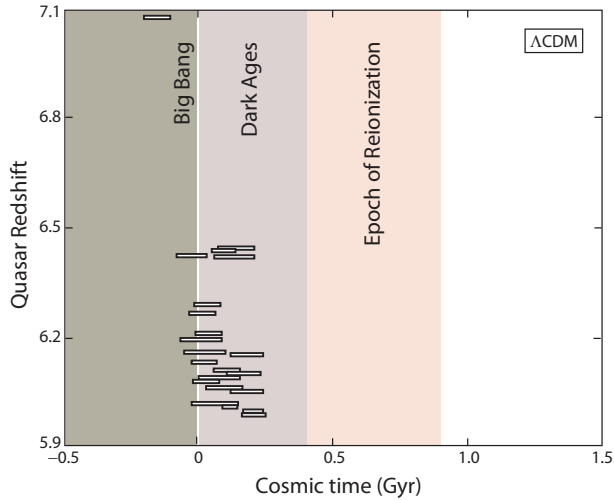} }}
\figcaption{Cosmic time $t$ (horizontal axis) at which the seed black hole, with a mass in the 
range $5-20\;M_\odot$, would have had to appear in order to produce the corresponding
quasar (from Table~1), observed at the redshift indicated on the vertical axis, via Eddington-limited accretion. 
The Epoch of Re-ionization (EoR) is limited roughly to the redshift range $\sim6-15$, which in $\Lambda$CDM 
corresponds to $t\sim 400-900$ Myr. The Dark Ages ($t\sim 0.4-400$ Myr in $\Lambda$CDM) is 
the period between recombination (at $\sim 0.4$ Myr in $\Lambda$CDM) and the onset of re-ionization.} 
\end{figure}

Insofar as accounting for the high-$z$ quasars is concerned, the essential
feature of the $R_{\rm h}=ct$ Universe that distinguishes it from $\Lambda$CDM is
its expansion factor, $a(t)\propto t$. Therefore, the cosmological redshift is given as
\begin{equation}
1+z={t_0\over t_e}\;,
\end{equation}
where $t_0$ is the current age of the Universe and $t_e$ is the cosmic time at which
the light with redshift $z$ was emitted. In addition, the gravitational horizon $R_{\rm h}$
is equivalent to the Hubble radius $c/H(t)$, and therefore one has $t_0={1/ H_0}$.
These equations are sufficient for us to produce a diagram like that shown in figure~1,
except this time for the $R_{\rm h}=ct$ Universe, and this is illustrated in figure~2. Note
that the EoR redshift range $z\sim 6-15$ here corresponds to the cosmic time $t\sim 830-
1,890$ Myr, and so the Dark Ages did not end until $\sim 830$ Myr after the big bang.

The most striking feature of this diagram, however, is that not only did all the high-$z$
quasars have time to grow after the big bang, but that all of the $5-20\;M_\odot$ seeds
that grew into these sources formed {\it after} the EoR had started. This is exactly
what one would have expected if these seeds were produced from Pop II
and III supernovae, which presumably occurred after these stars had time to evolve
and begin re-ionizing the Universe. To be clear, what these results are telling
us is that in the $R_{\rm h}=ct$ Universe, all of the high-$z$ quasars listed in Table 1
started their growth at redshift $z\la 15$, and developed into $M\sim 10^9\;M_\odot$
supermassive black holes by redshift $z\ga 6$. And all of this happened with standard
astrophysical principles, as we know them. There was no need for anomalously high
accretion rates, nor the exotic creation of very large seed black holes.

\begin{figure}[hp]
{\centerline{\epsscale{1.1} \plotone{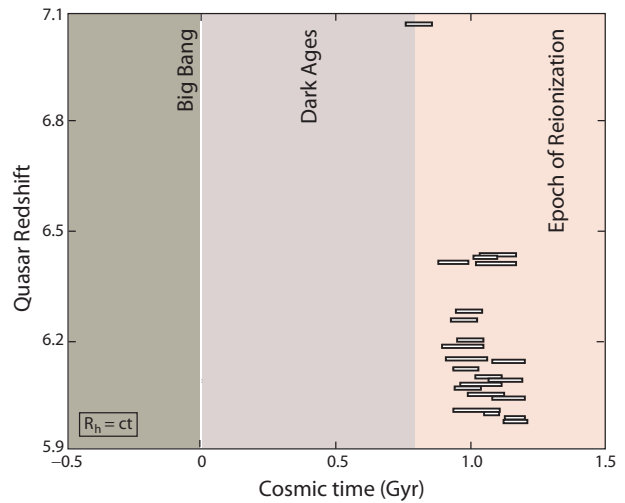} }}
\figcaption{Same as figure~1, except here for the $R_{\rm h}=ct$ Universe. In this case,
the EoR corresponds to $t\sim 800-1,900$ Myr, and the Dark Ages extend up to $\sim 800$ Myr.
Notice the close correspondence between the EoR and the appearance of the $5-20\;M_\odot$ 
seed black holes that formed the quasars we see at $z\ga 6$.} 
\end{figure}

\section{Conclusions}
It is difficult to understand how $\sim$$10^9\;M_\odot$ black holes appeared so
quickly after the big bang without invoking non-standard accretion physics
and the formation of massive seeds, both of which are not seen in the local Universe. 
The more recent data have compounded the problem by demonstrating that most (if
not all) of the high-$z$ quasars appear to be accreting at the Eddington limit. 

Within the framework of the standard model, it therefore appears that mergers
must have played a key role, not only in producing the observed correlations between 
black-hole mass and host bulge mass in the local Universe (Hirschmann et al. 2010), but 
perhaps also in assembling the $10^9\;M_\odot$ black holes seen at $z\sim 6-7$. 
However, the former outcome does not necessarily guarantee the latter. An
important finding of detailed merger simulations is that, irrespective of the initial
seed profile, the black-hole population always converges towards a Gaussian
distribution. Observations of supermassive black holes locally therefore allow
for quite some flexibility in the initial conditions.

Of course, this is quite useful for $\Lambda$CDM because the merger hypothesis
offers a viable mechanism by which the high-$z$ quasars could have formed
by $z\sim 6$. However, in order to comply with all of the available data, certain 
conditions must have prevailed at $z>>6$.  The simulations (see, e.g., Tanaka
\& Haiman 2009) show that $\sim 100\;M_\odot$ seeds must have started
forming by $z\sim 40$, well before the EoR, and must have stopped being
created, or accreted at severely diminished rates, by $z\sim 20-30$, in order
to not overproduce the mass density observed in lower-mass (a few $\times
10^5\;M_\odot$ to a few $\times 10^7\;M_\odot$) black holes. Without
this cutoff, these lower mass black holes would have been overproduced
by a factor of $\sim 10^2-10^3$.

These are all interesting possibilities and are permitted by the lack
of any direct observational constraints on black-hole mergers. In
principle, the {\it Laser Interferometer  Space Antenna} (LISA)
might have been able to detect such mergers, or place limits
on them, in the mass range $\sim (10^4-10^7)\;M_\odot /
(1+z)$ out to $z\sim 30$ (Tanaka \& Haiman 2009). But 
LISA may not become reality unless  a reprogrammed mission
can eventually be approved by ESA. Unfortunately,
the question of whether mergers could have assembled
the high-$z$ quasars in under $800$ Myr since the big
bang may therefore not be resolved any time soon.

There is also the question of whether the formation of
black-hole seeds in $\Lambda$CDM would be consistent
with our improved understanding of the EoR. It appears that
seeds would have formed quite early in the Universe's history,
well before re-ionization started at $z\sim 15$. Additional
work would have to be carried out to demonstrate that a
sufficient number of seeds could have been produced by
Population III stars shortly after the big bang, without any
evidence of re-ionization appearing until much later, past the 
point where the creation of seeds stopped (at $z\sim 20-30$).
It may not be possible to completely reconcile these opposing
trends. 

In this paper, we have suggested that the disparity between 
the predictions of $\Lambda$CDM and the observations is
further evidence against this being the correct cosmology. 
We have shown that the high-$z$ quasar data may instead 
be interpreted much more sensibly in the context of the 
$R_{\rm h}=ct$ Universe, for which standard $5-20\;M_\odot$ seeds
forming after re-ionization had begun (at $z\la 15$) could have grown to $\ga 10^9\;M_\odot$
supermassive black holes by redshift $z\ga 6$, merely by accreting at the observed
Eddington rate. Together with similarly compelling evidence from the CMB and the
large-scale matter distribution, this comparison suggests that $R_{\rm  h}=ct$ is
the correct cosmology and that $\Lambda$CDM is, at best, simply an empirically
derived approximation to it, that nonetheless fails at high redshifts.

\acknowledgments
I am grateful for helpful and enjoyable discussions with Xiaoui Fan and Andrew Shevchuk.
I am also grateful to Amherst College for its support through a John Woodruff Simpson Lectureship.
This work was partially supported by grant 2012T1J0011 from The Chinese Academy of Sciences Visiting 
Professorships for Senior International Scientists, and grant GDJ20120491013 from the Chinese
State Administration of Foreign Experts Affairs.

\end{document}